\documentstyle[preprint,revtex]{aps}

\tightenlines

\begin{document}

\preprint{TRIUMF TRI-PP-92-98}
\preprint{\ }
\preprint{October 1992}

\draft

\begin{title}
The Savvidy ``ferromagnetic vacuum'' \\
in three-dimensional lattice gauge theory
\end{title}

\author{Howard D. Trottier\cite{HDT} and R. M. Woloshyn}

\begin{instit}
TRIUMF, 4004 Wesbrook Mall, Vancouver, B.C., Canada V6T 2A3
\end{instit}

\begin{abstract}
The vacuum effective potential of three-dimensional
SU(2) lattice gauge theory in an applied
color-magnetic field is computed over a wide range of
field strengths. The background field is induced by an external
current, as in continuum field theory. Scaling and finite
volume effects are analyzed systematically.
The first evidence from lattice simulations is obtained
of the existence of a nontrivial minimum in the effective
potential. This supports a ``ferromagnetic'' picture of gluon
condensation, proposed by Savvidy on the basis of a
one-loop calculation in (3+1)-dimensional QCD.
\end{abstract}

\pacs{1990 PACS number(s): }

\narrowtext

The vacuum structure of quantum chromodynamics is known
to play a fundamental role in strong interaction
dynamics \cite{VacReview}, yet a clear understanding
of the physics underlying such basic properties of
the vacuum as the gluon condensate is lacking.
An attractive physical picture of gluon condensation
was proposed more than ten years ago by Savvidy,
who calculated the one-loop effective potential of
the QCD vacuum in response to an applied color-magnetic field
\cite{Savvidy}. The absolute minimum of the
effective potential was found to lie at a nonzero value of the
applied field, suggesting that the QCD vacuum behaves
like a ferromagnet, with a condensate in the color-field
arising spontaneously. This result was further analyzed by
Matinyan and Savvidy \cite{MatSav}, and was obtained
independently by Pagels and Tomboulis \cite{PagTom}.

However, the one-loop calculation is not adequate
{\it a priori\/} to support the ferromagnetic picture of
gluon condensation, since perturbation theory becomes
untrustworthy near the minimum of the effective potential.
Moreover it was discovered by Nielsen and Olesen
that a constant color-magnetic field generates
unstable modes at the one-loop level \cite{NielsenOlesen}.
Nevertheless, further analysis by the
Copenhagen group \cite{Copenhagen} suggested that the
instability is removed by a nonperturbative
structure that leaves the nontrivial minimum of the one-loop
effective potential intact. [This approach was challenged more
recently in Ref.\ \cite{Maiani} however, where it was argued
that the external field problem is fully nonperturbative,
even in the region of large fields.]
{}From another line of argument, Adler reached
the conclusion that the existence of a truly
nonperturbative minimum in the effective potential
can be established from the leading-order
calculation \cite{Adler}. An attractive model of hadronic
structure including the automatic formation of
flux-tubes and bag-like excitations has also been
developed based on the occurrence of a nontrivial minimum
in the effective potential \cite{Adler,OurQq}.

In order to truly substantiate the ferromagnetic
picture of the QCD vacuum, a genuine nonperturbative
calculation is required. In principle, the
effective potential is well suited to analysis
in lattice gauge theory. In fact, two groups have
recently attempted such a lattice calculation
in pure gauge SU(2) \cite{AmbLat,CeaLat}.
Unfortunately, the qualitative features of the
effective potential obtained by the two groups
are in sharp disagreement.
They acknowledge significant nonscaling and finite volume
effects in their calculations. Moreover, they were unable to
draw definitive conclusions about the behavior of the
effective potential at small fields, the region
of greatest interest.

We also note that very different algorithms for the
introduction of an external field on the lattice
were used in Refs.\ \cite{AmbLat} and \cite{CeaLat},
and we think that there are significant drawbacks to
both methods. In Ref.\ \cite{AmbLat} a link variable is introduced
to account for the external field; the external link
is aperiodic (corresponding to the usual external
vector potential in the continuum theory), while periodic
boundary conditions for the ``dynamical'' links are used.
We think that this procedure leads to undesired interaction
terms in the action on the lattice boundary.
In Ref.\ \cite{CeaLat} on the other hand
an Abelian projection technique is used to construct the
lattice action, and the results for the effective potential
turn out to depend strongly on the choice of operator
that is diagonalized.

In this Letter we present a calculation of the
effective potential in three-dimensional pure-gauge
lattice SU(2). Our rationale for doing the lattice
calculation in three dimensions is two-fold. First,
one of us has recently demonstrated that the
one-loop effective potential in 2+1 dimensions
has the same qualitative features as in
3+1 dimensions \cite{HDT3D}. In particular,
the effective potential has its absolute minimum
at a nonzero value of the applied color-magnetic
field in both theories. Thus the Savvidy
``ferromagnetic vacuum'' is found to
occur at the one-loop level in both
(2+1)- and (3+1)-dimensional QCD (the limitations of the
one-loop calculation are of course the same in both cases).
Secondly, one can more readily perform a thorough check of
scaling and finite volume effects in a three dimensional
lattice calculation that can be achieved (with the same
computing power) for the corresponding four-dimensional system.

We have also developed an algorithm for the introduction of a
background field in nonAbelian lattice gauge theory (in an
arbitrary number of dimensions) that is free of the
drawbacks in the methods used in Refs.\ \cite{AmbLat,CeaLat},
outlined above. Our algorithm is a straightforward
transcription of the procedure that is used in continuum field
theory to induce a background field by a coupling to an
external current. This approach has previously been considered
in the context of the lattice Abelian Higgs model by
Damgaard and Heller \cite{Heller}.

In the continuum U(1) theory, the (Euclidean) action for a
coupling of the gauge field to an external current is
\begin{equation}
   {\cal S}_{\rm U(1)}
   = \int dx \left[ {\case1/4} F_{\mu\nu}^2
   + j^{\rm ext}_\nu A_\nu \right] ,
\label{SU1j}
\end{equation}
where for our purposes the external current has the form
\begin{equation}
   j^{\rm ext}_\nu(x) = \partial_\mu F^{\rm ext}_{\mu\nu}(x) .
\label{jU1}
\end{equation}
If $F^{\rm ext}_{\mu\nu}(x)$ is constant within
a finite region, then $j^{\rm ext}_\mu(x)$ describes a
solenoidal current which circulates around the boundary.
An integration by parts in Eq.\ (\ref{SU1j}) leads to
\begin{equation}
   {\cal S}_{\rm U(1)} = \int dx \left[
     {\case1/4} F_{\mu\nu}^2
   - {\case1/2} F_{\mu\nu} F^{\rm ext}_{\mu\nu} \right] ,
\label{SU1}
\end{equation}
where now the region of support of the external field
is unrestricted. The path integral in this trivial theory
leads to a ``classical'' behavior for expectation values
such as $\langle F_{\mu\nu} \rangle = F^{\rm ext}_{\mu\nu}$.

An extension of the above procedure to nonAbelian
theories was proposed in Ref.\ \cite{CeaLat}. The action is
\begin{equation}
   {\cal S}_{\rm SU(N)}
   = \int dx \left[ {\case1/4} { F^a_{\mu\nu} }^2
   + j^{{\rm ext},a}_\nu A^a_\nu \right] ,
\label{Sj}
\end{equation}
where the current is a covariant generalization
of Eq.\ (\ref{jU1})
\begin{equation}
   j^{{\rm ext},a}_\nu(x) =
   {\cal D}^{ab}_\mu (A^{\rm ext})
    F^{{\rm ext},b}_{\mu\nu}(x) ,
\label{j}
\end{equation}
with ${\cal D}^{ab}_\mu (A) \equiv
\partial_\mu \delta^{ab} - g f^{abc} A^c_\mu$
the usual covariant derivative.
$F^{{\rm ext},a}_{\mu\nu}$ is the nonAbelian field strength
constructed out of $A^{{\rm ext},a}_\mu$.
Following the one-loop calculations of the effective potential
\cite{Savvidy,NielsenOlesen,Copenhagen,HDT3D}, we consider
an external field with a fixed direction in gauge space
\begin{equation}
   A^{{\rm ext},a}_\mu = A^{\rm ext}_\mu \delta^{a3}
   \quad \Rightarrow \quad
   F^{{\rm ext},a}_{\mu\nu} = F^{\rm ext}_{\mu\nu} \delta^{a3} .
\label{Aexta}
\end{equation}
An integration by parts in Eq.\ (\ref{Sj}) then yields
\begin{equation}
   {\cal S}_{\rm SU(N)}  = \int dx \left[
     {\case1/4} F^a_{\mu\nu} {}^2
   - {\case1/2}
     {}^\partial \! F^{a=3}_{\mu\nu} \,
     F^{{\rm ext}}_{\mu\nu}  \right] ,
\label{S}
\end{equation}
where
\begin{equation}
   {}^\partial \! F^{a=3}_{\mu\nu} \equiv
      \partial_\mu A^{a=3}_\nu
    - \partial_\nu A^{a=3}_\mu .
\label{dF}
\end{equation}
Notice that the interaction term in Eq.\ (\ref{S})
only contains the derivative part ${}^\partial \! F$ of the
field strength for the ``dynamical'' gauge fields
$A^a_\mu$ (compare ${}^\partial \! F$ with the full field
strength tensor $F^a_{\mu\nu} =
\partial_\mu A^a_\nu - \partial_\nu A^a_\mu
- g f^{abc} A^b_\mu A^c_\nu$). If the
full field strength appeared in the interaction term,
then Eq.\ (\ref{S}) would describe the same ``trivial'' physics
as the U(1) action of Eq.\ (\ref{SU1}). This difference
between the two actions expresses the fact that the
``charge'' carried by nonAbelian gauge fields
drives the different physics.

Equations (\ref{Sj}) and (\ref{S})
represent a sensible way to introduce an external field
in the nonAbelian theory since they lead, at the classical
level, to an ``induced'' gauge field that is equal to the
applied field; moreover, at the one-loop level, this
action leads to the Savvidy effective potential of interest.

We now consider the lattice transcriptions of the above
continuum theories. We specialize to three dimensions.
The lattice equivalent of the U(1) action Eq.\ (\ref{SU1})
is \cite{Heller}
\begin{mathletters}
\begin{equation}
   {\cal S}_{\rm U(1)} =
   {\cal S}_W
 - {\case1/2} \sum_{\rm sites}
   \widehat F_{\mu\nu} \widehat F^{\rm ext}_{\mu\nu} ,
\label{SU1L}
\end{equation}
where ${\cal S}_W$ is Wilson's plaquette action,
the external field in lattice units
$\widehat F^{\rm ext}$ is given by
\begin{equation}
   \widehat F^{\rm ext}_{\mu\nu} =
   a^{3/2} F^{\rm ext}_{\mu\nu} ,
\label{FextU1L}
\end{equation}
and the ``dynamical'' field
strength $\widehat F_{\mu\nu}$ is given as usual in terms
of the imaginary part of the plaquette $U_{\mu\nu}$:
\begin{equation}
   \widehat F_{\mu\nu} =
   \sqrt\beta \, {\rm Im} \, U_{\mu\nu} ,
   \quad \beta \equiv {1 \over e^2 a}
\label{FU1L}
\end{equation}
\end{mathletters}
($a$ the lattice spacing). In the SU(2) lattice theory:
\begin{mathletters}
\begin{equation}
   {\cal S}_{\rm SU(2)} =
   {\cal S}_W
 - {\case1/2} \sum_{\rm sites}
   {}^\partial \! \widehat F^{a=3}_{\mu\nu} \,
   \widehat F^{{\rm ext}}_{\mu\nu} ,
\label{SL}
\end{equation}
where ${\cal S}_W$ is Wilson's plaquette action for SU(2), and
\begin{equation}
   \widehat F^{{\rm ext}}_{\mu\nu} =
   \left( {4 \over \beta} \right)^{3/2} {1 \over g^3} \,
   F^{{\rm ext}}_{\mu\nu} ,
   \quad \beta \equiv {4 \over g^2 a} .
\label{FextL}
\end{equation}
To extract the derivative part ${}^\partial \! \widehat F$
of the ``dynamical'' field strength on the lattice, we
explicitly compute the commutator of
any two links $U_\mu$, $U_\nu$ which span the plaquette:
\begin{equation}
   {}^\partial \! \widehat F^{a=3}_{\mu\nu} =
   -i \sqrt{\beta} \,
   {\rm Tr} \, \left\{
   \left(  U_{\mu\nu} - [ U_\mu, U_\nu ] \right)
   {\case1/2} \sigma^3 \right\} ,
\label{dFL}
\end{equation}
\end{mathletters}
where $\sigma^3$ is the third Pauli matrix. The
connection between the right hand side of Eq.\ (\ref{dFL})
and ${}^\partial \! F$ in the continuum limit
follows from the identification of the link variables in
this limit as $U_\mu \equiv \exp(i a g A^a_\mu \sigma^a/2)$;
the above lattice action reduces to the continuum action
Eq.\ (\ref{S}), up to corrections of $O(a)$, which is the same
order of accuracy as the continuum limit of the Wilson action.
[The coupling to the external current which leads to
Eq.\ (\ref{SL}) breaks the local SU(2) gauge invariance of
the Wilson action to a local U(1) symmetry. This implies, for
example, that ${\rm Tr} ( U_{\mu\nu} \sigma^3 )$ is an
invariant quantity in the theory defined by Eq.\ (\ref{SL}).]
Our construction of the lattice action using
Eq.\ (\ref{dFL}) differs from the procedure followed
in Ref.\ \cite{CeaLat}, where ${}^\partial \! \widehat F$
was defined by an Abelian projection (the results for the
effective potential obtained in Ref. \cite{CeaLat} depend
strongly on the choice of operator that is diagonalized).

In our calculations we take the field strength tensor for
the external magnetic field (a scalar in three dimensions)
to have nonvanishing components only in the (1,2) plane:
\begin{equation}
   F^{{\rm ext}}_{\mu\nu} = H \left(
   \delta_{\mu 1}\delta_{\nu 2}
 - \delta_{\nu 1}\delta_{\mu 2} \right) .
\label{F12}
\end{equation}

An important aspect of our lattice simulation
is our use of {\it free boundary conditions\/} for
the ``dynamical'' gauge links. That is, we integrate over
all links on the boundaries of the lattice \cite{Free}.
This is motivated by the fact that the gauge field for a
uniform magnetic field in the usual continuum theory is
not periodic [cf.\ $A^{\rm ext}_\mu = x H \delta_{\mu 2}$,
which generates Eq.\ (\ref{F12})].
In a perturbative calculation moreover, the
continuum external potential induces quantum fluctuations
which are also aperiodic. Free boundary conditions seem most
appropriate to describe the corresponding physics on the lattice.
Free boundary conditions also serve to eliminate long-lived
metastable states that occur in the U(1) theory with
periodic boundary conditions, corresponding to closed Dirac
strings winding through the (dual) lattice \cite{U1string}.
In a simulation of the U(1) external field problem on a
lattice with periodic boundary conditions, it is necessary to
use a global updating procedure to eliminate
these strings \cite{Heller}.

We compute the energy $\widehat{\cal E}(H)$ in the ``induced''
gauge fields in lattice units according to
(Refs.\ \cite{Energy,AmbLat,CeaLat})
\begin{mathletters}
\begin{equation}
   \widehat{\cal E}(H) \equiv \beta
   \left[  \langle \Box_M \rangle (H)
         - \langle \Box_E \rangle (H) \right] ,
\label{hatE}
\end{equation}
where
\begin{eqnarray}
   & & \langle \Box_M \rangle \equiv {1 \over L^3}
       \sum_{\rm sites}
       \left[ 1 - {\case1/2} {\rm Tr} \, U_{12} \right] ,
\nonumber \\
   & & \langle \Box_E \rangle \equiv {1 \over L^3}
       \sum_{\rm sites}
       \left[ 2 - {\case1/2} {\rm Tr}
       \left( U_{13} + U_{23} \right) \right] ,
\label{boxME}
\end{eqnarray}
\end{mathletters}
and $L$ is the length of a side of the lattice.
To get the physical energy ${\cal E}_{\rm phys}$
we perform a vacuum subtraction, and convert to units of
the coupling constant $g$ (which in three dimensions
has units of (mass)$^{1/2}$). For SU(2):
\begin{equation}
   {\cal E}_{\rm phys} = g^6
   \left( {\case1/4} \beta \right)^3
   \left[ \widehat{\cal E}(H) - \widehat{\cal E}(0) \right] .
\label{Ephys}
\end{equation}

We now present results of our simulations of the U(1) and
SU(2) theories. We have performed an extensive set of
calculations over a wide range of values of $H$, $\beta$,
and lattice sizes. We find that relatively modest statistics
(similar to those employed in the previous
four-dimensional studies of Refs.\ \cite{AmbLat,CeaLat})
are sufficient to identify the main features of the
effective potential. We use a 10-hit Metropolis algorithm.
Our main results in the SU(2) case were obtained on
a $32^3$ lattice, using 1000 sweeps to thermalize
at each value of $H$ and $\beta$, followed by
5000 sweeps, keeping only every fourth configuration
for data (this results in an integrated
autocorrelation time in the energy data which in
most cases satisfies $\tau_{\rm int} \alt 0.6$).
To check finite volume effects, we have also done
some calculations in the SU(2) case on a $16^3$
lattice with an eightfold increase in statistics.

Our U(1) calculations provide a useful check of our
SU(2) code --- we simply identify the link variables
$U_\mu^{a=1,2} \equiv 0$ to do the U(1) simulation using our
SU(2) code. Figure \ref{FigU1} shows results obtained
on a $16^3$ lattice (using 1000 sweeps to thermalize at
each value of $H$, followed by 10000 sweeps, keeping only every
fourth configuration for data). The agreement with the expected
classical behavior in the energy and in the induced
field is very good, to within about 5\% at $\beta=10$
(the difference from the classical values is due mainly to
finite size effects, and decreases with increasing $\beta$).

Figures \ref{FigEbig} and \ref{FigEsmall} show the
energy in the SU(2) theory on a $32^3$ lattice as a
function of $H$, for various values of $\beta$. We note that
at large $H$ one must run at comparatively large values
of $\beta$ in order to approach the scaling limit,
as evident in Fig.\ \ref{FigEbig}.
[Roughly speaking, we may expect some components
of the ``induced'' field to be of $O(H)$, corresponding
to a plaquette angle of $O(a^2 g H)$, which
should remain $\ll 1$ in order to approach the
continuum limit (cf.\ Ref.\ \cite{AmbLat}).]
Hence there may be appreciable finite volume effects at
the largest fields in our data.

In the region $H/g^3 \alt 1$ on the other hand,
our data shows both good scaling behavior
(cf.\ Fig.\ \ref{FigEsmall}), and negligible
finite volume effects, as shown in Fig.\ \ref{FigEvol},
where we compare our results from $16^3$ and $32^3$
lattices. Our data at $\beta = 7$ and 10 show clear
evidence of a minimum in the energy. The data at $\beta=7$
lie about eight standard deviations below zero in a
region of $H/g^3$ near 1.
[The existence of a minimum is somewhat
ambiguous in the case of our data at $\beta=12$, since
the errors in ${\cal E}_{\rm phys}$ with the present
level of statistics are large compared with its magnitude
for $\beta=12$ and $H/g^3 \alt 2$ (finite volume effects in
these data are also significant, cf.\ Fig.\ \ref{FigEvol}).
However, the data at $\beta=12$ are
generally consistent within errors with the energy
obtained at $\beta=7$ and 10.]
The energy is generally more than an order of magnitude
smaller than the classical result ${\case1/2} H^2$ over
the whole range of $H$ that we have studied.

In summary, we have developed a method for the introduction
of an external field in nonAbelian lattice gauge theories that
follows from a coupling to an external current, as used in
continuum field theories. We have performed a systematic
study of scaling and finite volume effects, and we have
obtained results for the effective potential over a
wide range of field strengths. We have obtained the first
evidence from lattice simulations of the existence of a
nontrivial minimum in the effective potential for
a background color-magnetic field in nonAbelian
gauge theory. Our results support the Savvidy
``ferromagnetic'' picture of gluon condensation
in the QCD vacuum. In future work, we will use the
methods developed here to compute the effective
potential in four-dimensional lattice SU(2).
It should also be possible to look for the
nonperturbative vortex structure associated with
the ``would-be'' Nielsen-Olesen unstable
modes \cite{NielsenOlesen,Copenhagen}. The
three-dimensional gauge theory studied here should provide
a very convenient system for such a study.

HDT thanks Jan Ambj\o rn and the Niels Bohr Institute
for their hospitality during a stay which provided
for fruitful discussion of this work. This work was
supported in part by the Natural Sciences and Engineering
Research Council of Canada.

\figure{Effective potential and expectation value
of the induced field strength for three-dimensional
U(1) lattice gauge theory, as functions of the applied
magnetic field strength $H$. These results are from
a $16^3$ lattice at $\beta=10$.\label{FigU1}}

\figure{Effective potential for three-dimensional SU(2)
lattice gauge theory, as a function of the applied
color-magnetic field strength $H$, for various
values of $\beta$. These results were
obtained on a $32^3$ lattice.\label{FigEbig}}

\figure{Effective potential for three-dimensional
lattice SU(2), as in Fig.\ \ref{FigEbig}, here shown
on an expanded scale for ``small'' field strengths
$H/g^3 \leq 2$.\label{FigEsmall}}

\figure{Effective potential for three-dimensional lattice
SU(2) on two different lattice sizes. Data points for
$H/g^2 \leq 1$ were calculated at $\beta =7$, while the points
at larger $H$ were taken at $\beta = 12$.\label{FigEvol}}

\end{document}